\documentclass{elsart}
\usepackage{graphicx}
\begin{document}

\begin{frontmatter}
\title{Stochastic Multiplicative Processes for Financial Markets}
\author{Zhi-Feng Huang\thanksref{newadd1}}
\address{Department of Physics, University of Toronto,
Toronto, ON, Canada M5S 1A7}
\author{Sorin Solomon\thanksref{newadd2}}
\address{Racah Institute of Physics, The Hebrew University, 
Jerusalem 91904, Israel}
\thanks[newadd1]{{\em E-mail address}: zfh@physics.utoronto.ca}
\thanks[newadd2]{{\em E-mail address}: sorin@vms.huji.ac.il}
\date{}
\maketitle
\begin{abstract}
We study a stochastic multiplicative system composed of finite
asynchronous elements to describe the wealth evolution in financial 
markets. We find that the wealth fluctuations or returns of this 
system can be described by a walk with correlated step sizes 
obeying truncated L\'{e}vy-like distribution, and the 
cross-correlation between relative updated wealths is the origin 
of the nontrivial properties of returns, including the power law 
distribution with exponent outside the stable L\'{e}vy regime and 
the long-range persistence of volatility correlations.

\noindent PACS: 05.40.-a; 87.23.Ge; 05.70.Ln
\end{abstract}

\begin{keyword}
Multiplicative processes; Power law; Volatility correlations.
\end{keyword}
\end{frontmatter}

\section{Introduction}

Multiplicative processes have been well studied in different
contexts and widely applied to various research fields, such
as the biological, social, and economic systems
\cite{Kesten73,Redner90,Solomon96,Solomon98,Sornette,Marsili98}.
One of the major interests in these processes is the generation
of power laws \cite{Kesten73,Solomon96,Solomon98,Sornette,Bouchaud00},
which have been observed in many natural domains and indicate
the properties of scale invariance and universality. In fact,
the investigations of power law behaviors in systems with stochastic
multiplicative dynamics can be traced back to decades ago, and 
the underlying physical mechanisms are still to be understood.

For the applications in financial problems, the assumption of
multiplicative property for the individual capital investments
$w_i$ (the index $i=1,...,N$ may correspond to various 
investors/traders or companies (stocks) in the market) leads to
a term
\begin{equation}
w_i(t+1) \sim \lambda(t) w_i(t)
\label{eqMul}
\end{equation}
appearing in the dynamics of wealth/capital evolution. That is,
the individual capital $w_i$ at time $t+1$ is proportional to
the invested capital itself. The random factor $\lambda(t)$
reflects the relative gain/loss incurred by individuals between 
time $t$ and $t+1$, and is chosen from a probability distribution
$\Pi(\lambda)$.

The Pareto power law distribution \cite{Pareto} of individual 
wealths $w$
\begin{equation}
P(w)\sim w^{-1-\alpha_w},
\label{eqPw}
\end{equation}
has been found in many of the previous studies of multiplicative
processes \cite{Kesten73,Solomon96,Solomon98,Sornette,Marsili98,Bouchaud00},
which all follow the above dynamics (\ref{eqMul}) (plus some additional
crucial dynamical elements that we will discuss later),
and been verified repeatedly in the last hundred years
in most of the capitalistic societies, e.g., for individual income 
and wealth \cite{Pareto,Atkinson78}, size of business firms
\cite{Ijiri77}, etc. In spite of the significant fluctuations in the 
total wealth (with occasional spectacular booms and crashes),
the exponent $\alpha_w$ was observed typically in the range 
$1<\alpha_w <2$, i.e., within the stable L\'{e}vy regime \cite{Levy37}.

Besides the above distribution property of individual wealth, 
we are more interested in studying the relative fluctuations of
the total wealth $W(t)=\sum_{i=1}^N w_i(t)$,
which may represent the capitalization 
(total market value) of a company, or the market price of a stock 
(when normalized to the total number of shares that stock has on 
the market), or the market index. The fluctuations of $W(t)$ over 
arbitrary periods of time $\tau$ define the market returns $r(\tau)$ 
during these time periods:
\begin{equation}
r(\tau) = \ln W(t+\tau) - \ln W(t),
\label{eqreturn}
\end{equation}
which have attracted much attention in recent financial studies,
in particular for their statistical properties 
\cite{Stanley-book,Bouchaud-book}. There are some 
empirical features generic for different financial markets, 
including: 

\begin{itemize}
\item The power-law decay of probability density of returns $r$
\begin{equation}
P(r)\sim r^{-1-\alpha},
\label{eqPr}
\end{equation}
in tail region, but with exponent $\alpha>2$ different from
that of wealth distribution (\ref{eqPw}), i.e.,
outside the L\'{e}vy regime \cite{Lux96,Gopikrishnan99,Plerou99}.
For larger $r$ values, $P(r)$ can be fitted by an exponential
in some observations \cite{Huang00,Skjeltorp00,Masoliver00};

\item The very short range (about a few minutes) correlation of 
returns and long-range persistence for correlation of square
or absolute value of returns (so-called volatility clustering)
\cite{Stanley-book,Bouchaud-book,Gopikrishnan99,Huang00}.
\end{itemize}
It is interesting to not only reproduce these stylized
facts through microscopic models, but also understand the
intrinsic mechanisms dominating the processes.

\section{Generalized Lotka-Volterra system}

The framework we use to describe the dynamics of financial
systems is the Generalized Lotka-Volterra (GLV) model proposed 
by one of us a few years ago \cite{Solomon96,Solomon98}:
\begin{equation}
w_i(t+1) = \lambda(t) w_i(t) + a \bar{w}(t)
+c(w_1,w_2,...,w_N,t) w_i(t),
\label{eqGLV}
\end{equation}
with the average wealth $\bar{w}(t)=W(t)/N$. The multiplicative
property at the individual level, i.e., Eq.\ (\ref{eqMul}), is 
expressed in the first r.h.s. term, and the second term 
represents the property of wealth at the social level,
which may correspond to the social security, subsidies, or
funded services, and can also be interpreted as arising from
the diffusion of wealth between agents (by services, taxes, etc.). 
This term is set to be proportional to the average wealth 
$\bar{w}$ with an important coupling parameter $a$,
supplying the correlation and coupling between investors or 
companies that is crucial for the distribution properties of 
wealth and wealth fluctuation shown below. Generally, the
last term $c w_i$ which corresponds to the competition
in the market, does not qualitatively affect the properties
of the model \cite{Solomon01}. For simplicity, we set it $0$ 
in the present study. Therefore, we have
\begin{equation}
w_i  (t+1) = \lambda (t) w_i (t) + a \bar{w} (t).
\label{eqwi}
\end{equation} 

Note that in this stochastic multiplicative system an 
asynchronous updating mechanism 
\cite{Solomon96,Solomon98,Biham-Malcai,Huang-Solomon01,Biham01}
is used, due to the fact that in a real market the capitals 
of different agents or companies are not updated simultaneously. 
Thus, at each discrete time $t$, one of the elements $i$ is 
chosen randomly and updated according to Eq.\ (\ref{eqwi}),
while the other $w_i$'s keep unchanged. Moreover, the random
factor $\lambda$ is taken in a rather narrow range around $1$,
since in practice the price returns for very small time
intervals are usually rather small.

\section{Power law behaviors and cut-off effect}
\label{sec3}

The system described by Eq.\ (\ref{eqwi}) does not approach a
steady state; instead the total wealth grows (or decays) 
exponentially with superimposed fluctuations, i.e.,
$W(t)\sim \exp(\kappa t/N)$ with $\kappa$ depending on $a$
as well as the mean and standard deviation of $\lambda$ 
distribution. Nevertheless, the instantaneous values of wealth 
(or the relative value $w_i(t)/W(t)$) fulfill the Pareto 
power-law distribution (\ref{eqPw}), with 
exponent $\alpha_w<2$ \cite{Biham-Malcai}, that is, inside
the stable L\'{e}vy regime.

More interesting results for this multiplicative system
are obtained by studying the wealth fluctuations (i.e., 
returns) $r(\tau)$ defined by Eq.\ (\ref{eqreturn}).
(Note that due to the growth effect of total wealth $W$
($\sim\exp(\kappa t/N)$) in this model, the return in
(\ref{eqreturn}) should be detrended, that is, change by a 
constant: $r(\tau) \rightarrow r(\tau)-\kappa\tau/N$.)
For a certain time interval $\tau$, the return $r(\tau)$
can be written as
$\sum_{s=t}^{t+\tau-1} \ln [(W(s+1)-W(s))/W(s)+1]$$=
\sum_s \ln [(\lambda(s)-1)w_i(s)/W(s)+a/N+1]$ according
to Eq.\ (\ref{eqwi}) \cite{note2}. Since the relative value
of wealth $w_i/W$ at each step is small
and generally $\lambda-1$ is in a narrow range around
$0$, one obtains that the properties of the (detrended) return 
$r(\tau,t)$ are the resultant of $\tau$ steps walk
\begin{equation}
R(\tau) = \sum\limits_{s=t}^{t+\tau-1} (\lambda(s)-1)w_i(s)/W(s),
\label{eqwalk}
\end{equation}
with step sizes governed by power law distribution (\ref{eqPw})
with cut-off effect.

For $\tau=1$ the distribution of returns differs from that of
the relative wealth $w_i(t)/W(t)$ only by a random factor
$\lambda(t)-1$, and then displays a similar power law behavior
(\ref{eqPr}) with exponent $\alpha\sim\alpha_w$ and a sharp
peak around $0$. As $\tau$ increases, the consequence of $\tau$
step walks described by Eq.\ (\ref{eqwalk}) is the smoothening 
of this tip.

Note that for any finite $N$ the distribution of $w_i(t)/W(t)$
is not a perfect power law even if one assumes that of $w_i$ to
be so, since this distribution is truncated from above:
$w_i/W<1$, shown as a bending in a 
log-log plot for very large values of $w_i$ \cite{Huang01}. 
We have proven \cite{Huang-Solomon01} that this 
upper cut-off effect leads to an exponential decay for extremely 
large values of returns and for very wide range of interval $\tau$.

Thus, the process described by Eq.\ (\ref{eqwalk}) looks like a 
truncated L\'{e}vy walk leading to similar power laws
for different $\tau$ and slow convergence to Gaussian 
\cite{Mantegna94}. However, our studies below indicate a more
complicated situation, that is, this process is different from 
the ordinary L\'{e}vy walk due to the coupling between step 
sizes.

\begin{figure}
\centerline{
\resizebox{0.75\textwidth}{!}{%
  \includegraphics{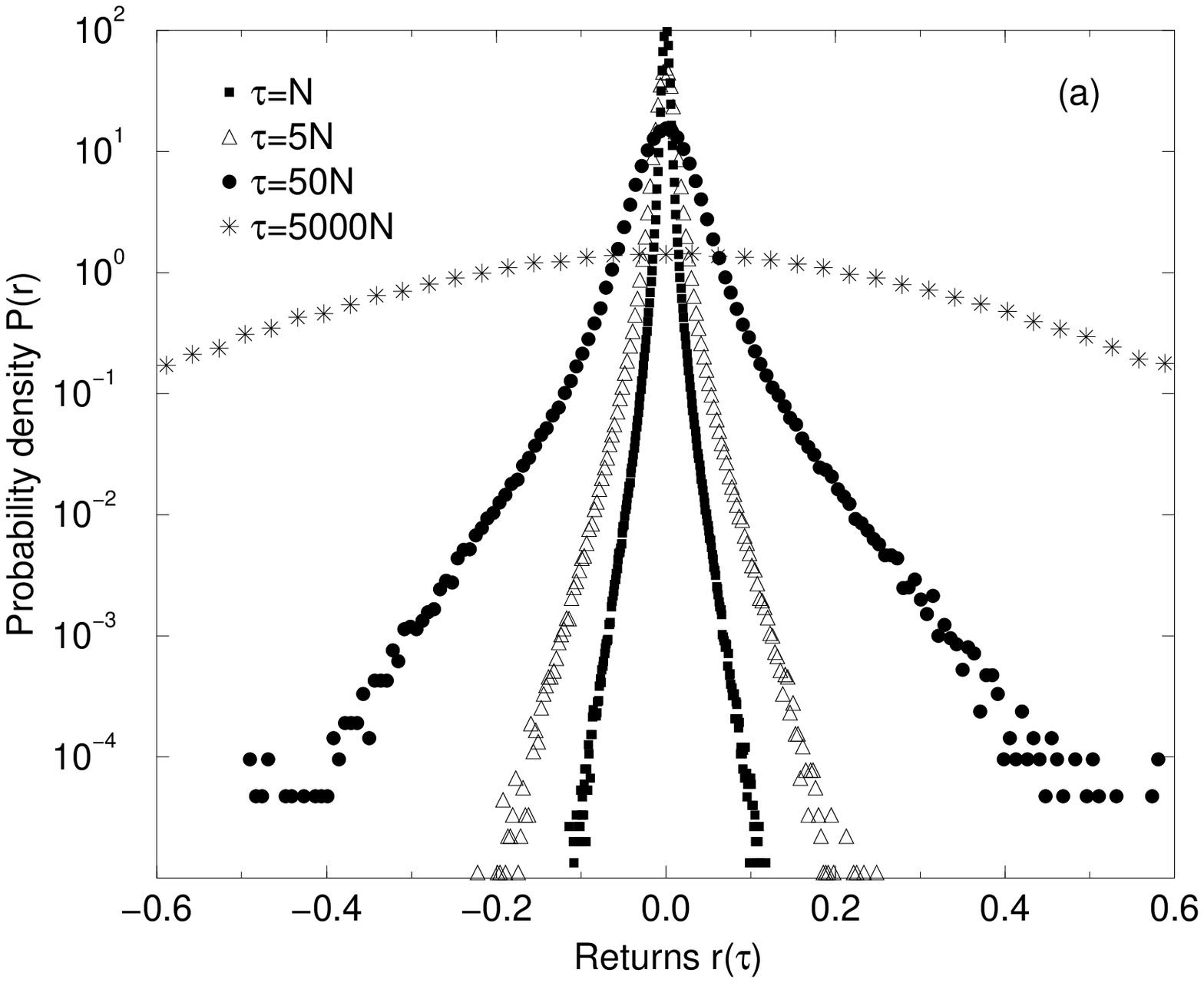}}}

\centerline{
\resizebox{0.75\textwidth}{!}{%
  \includegraphics{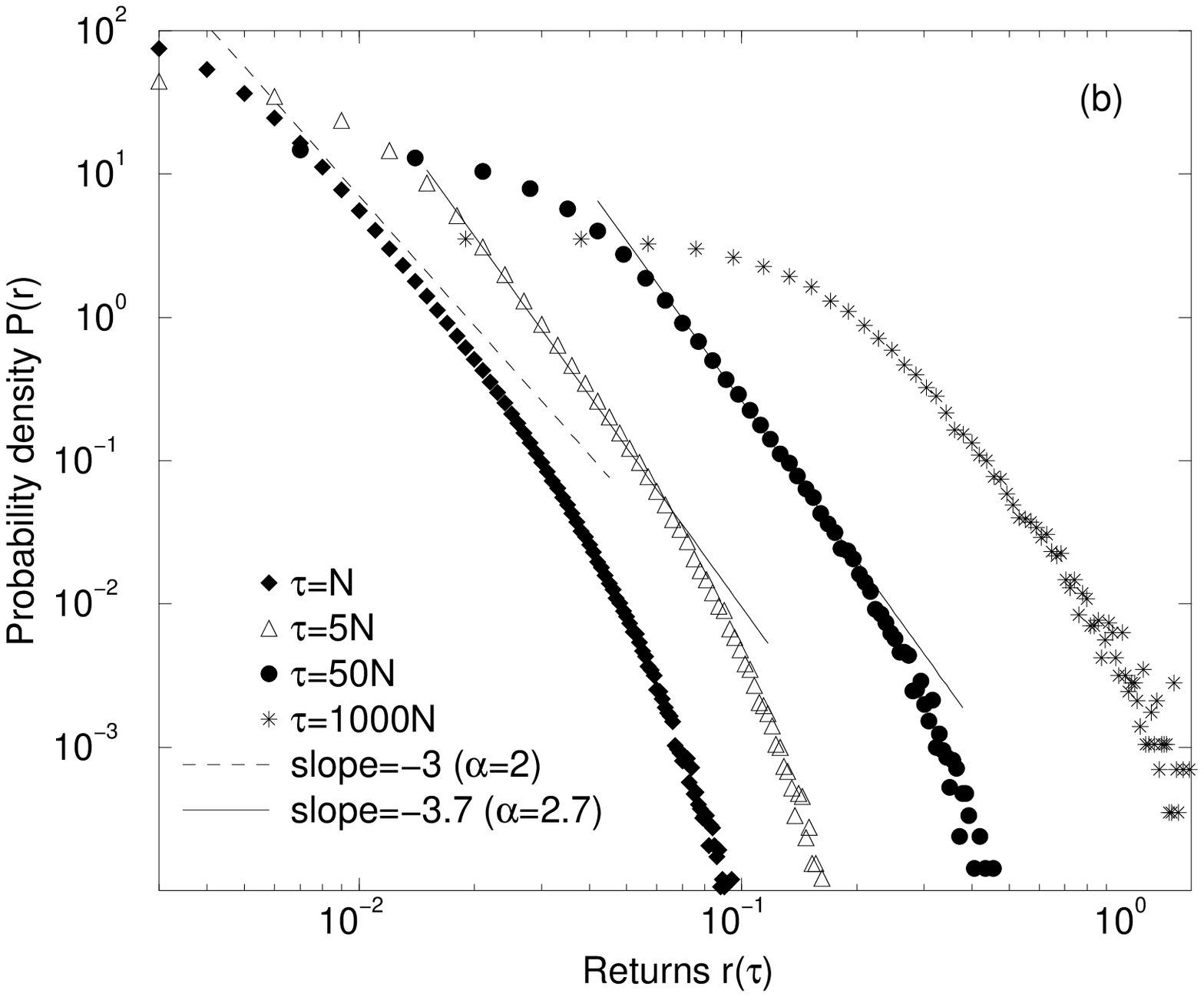}}}
\caption{The probability distribution of returns $r(\tau)$
defined by Eq.\ (\ref{eqreturn}), for $N=500$ and different time 
intervals $\tau$ from $N$ to $5000N$. $\lambda$ uniformly 
distributes between $0.95$ and $1.05$, and $a=0.0002$.
The results are averaged over $1500$ runs for $\tau\leq 1000N$ 
and $600$ runs for $\tau=5000N$.(a) Semi-log plot, where 
for the largest $\tau$ ($=5000N$) Eq.\ (\ref{eqwalk}) is used; 
(b) Log-log plot for positive returns.}
\label{figr500}
\end{figure}

\begin{figure}
\centerline{
\resizebox{0.75\textwidth}{!}{%
  \includegraphics{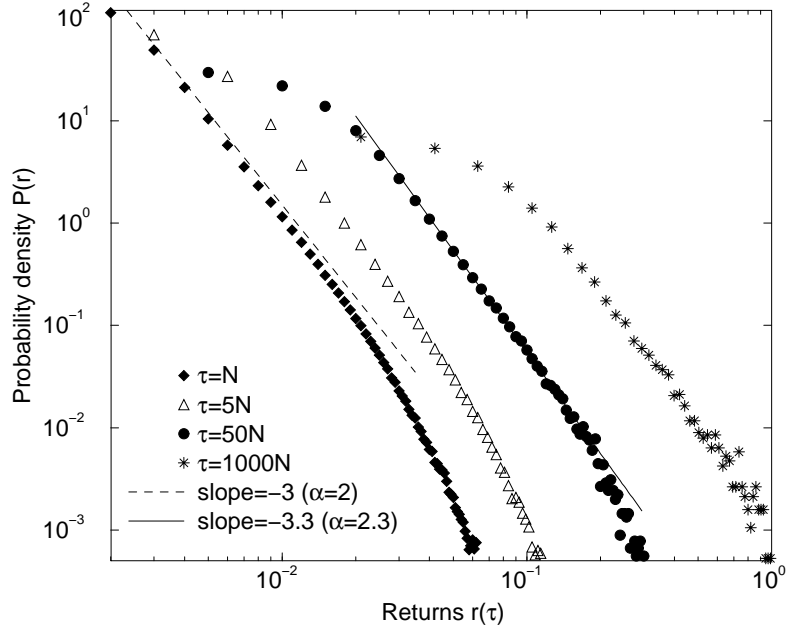}}}
\caption{Log-log plot of the probability distribution of
positive returns $r(\tau)$ (averaged over $100$ runs), for 
system size $N=5000$ and different time intervals $\tau=N$, 
$5N$, $50N$, and $1000N$.}
\label{figr5000}
\end{figure}

The numerical results of Eq.\ (\ref{eqwi}) for the distribution 
of returns $r(\tau)$ (\ref{eqreturn}) are shown in Figs.\ 
\ref{figr500} and \ref{figr5000}, where the random gain/loss 
factor $\lambda$ has an uniform distribution in the range 
$0.95 < \lambda < 1.05$ and the coupling parameter $a=0.0002$, 
with the corresponding power-law exponent of wealth distribution 
$\alpha_w$ about $1.5$ as in usual financial application. 
(In all the numerical simulations of this paper the results are
calculated after $t=10^5 N$ updatings, so that the returns series
are stationary after detrending the wealth growth factor.)
As expected according to the above analysis, the 
semi-log plots of distribution in Fig.\ \ref{figr500} (a) exhibit 
the smearing out of the sharp peak for small time intervals into a
dome-like shape for large one, and one obtains the exponential-type 
behavior for extremely large $r$ due to the cut-off effect of 
finite $N$ system, even for very large interval $\tau$, which has 
been observed in some empirical studies 
\cite{Huang00,Skjeltorp00,Masoliver00}.

The power-law scaling region described by Eq.\ (\ref{eqPr})
can be found in the log-log plots of the distribution (see 
Figs.\ \ref{figr500} (b) and \ref{figr5000}), and interestingly,
exponents much larger than that of the wealth distribution
(\ref{eqPw}) are obtained. For small interval $\tau$, 
the exponent $\alpha$ is within the stable L\'{e}vy regime, i.e., 
$\alpha_w < \alpha <2$, while for large intervals ($\tau>N$)
we obtain the result consistent with the recent empirical 
observations for both the stock index
(German DAX \cite{Lux96} and S\&P 500 \cite{Gopikrishnan99})
and individual stocks \cite{Plerou99}, that is, the effective 
exponent $\alpha>2$ outside the L\'{e}vy regime. Similar to the 
findings of real markets \cite{Lux96,Gopikrishnan99,Plerou99}, 
the exponent increases very slightly with the increase of the time 
interval, due to the very slow convergence to Gaussian, 
as shown in Figs.\ \ref{figr500} and \ref{figr5000}.

The extension of this power-law region with exponent $\alpha>2$
depends on the size $N$ of the multiplicative system. By 
comparing the results in Fig.\ \ref{figr500} for $N=500$ with 
that in Fig.\ \ref{figr5000} for $N=5000$, one can find that the 
range of the power-law scaling is longer for larger systems. 
This phenomenon is attributed to the exponential
truncation effect discussed above for finite $N$, which appears 
more obvious for smaller system. As observed in the log-log 
plots, the deviation and bend-down from a straight line occurs 
for large $r$, which has been found very recently in the Hang
Seng Index (HSI) of Hong Kong \cite{Huang00} if one skips
the data of first 20 minutes in daily opening. The similar effect 
of finite size $N$ has been found in some microscopic models,
in particular the Cont-Bouchaud percolation-type model 
\cite{Stauffer99}, where the $\alpha>2$ power-law exponent is 
obtained over an intermediate region which becomes infinitely 
long for infinite large market size and computer time, while 
the asymptotic return distribution is expected to be a 
(stretched) exponential decay in the tails.

However, different from the percolation model, the exponent
$\alpha$ for large interval $\tau$ decreases with increasing 
size $N$ in our system. As shown in Figs.\ \ref{figr500} (a)
and \ref{figr5000}, $\alpha \sim 2.7$ for $N=500$, and
$\alpha \sim 2.3$ for $N=5000$. Moreover, for $N=10^5$
(not shown here) $\alpha \sim 2$, which is the border of
the stable L\'{e}vy regime. Therefore, the property of
returns distribution is size-dependent, and one can obtain 
that the effective $N$ for real market is not very large 
according to the high empirical $\alpha$ value.

\begin{figure}
\centerline{
\resizebox{0.75\textwidth}{!}{%
  \includegraphics{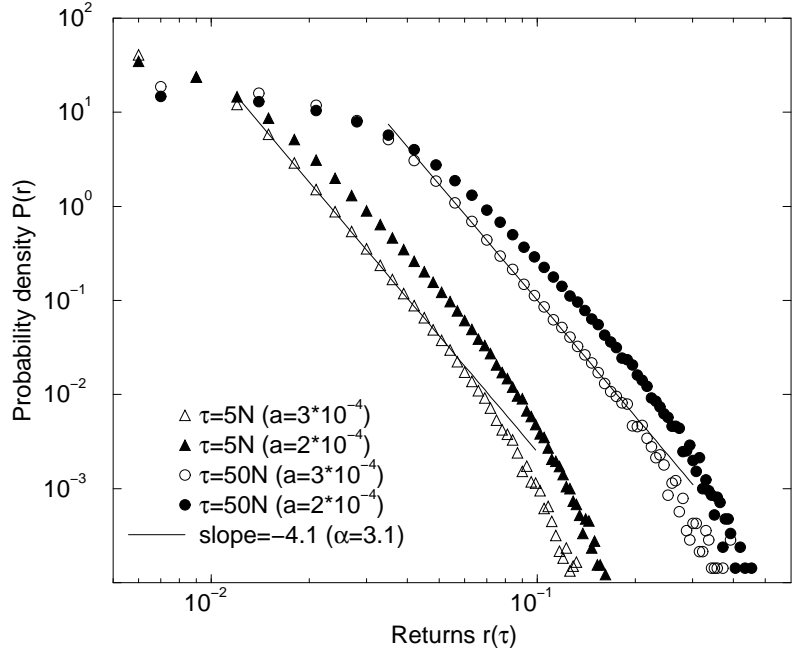}}}
\caption{Log-log plot of the probability distribution of
positive returns $r(\tau)$, for $N=500$ and different $a$ 
values $0.0003$ and $0.0002$. Other parameters are the same as 
Fig.\ \ref{figr500}.}
\label{figra23}
\end{figure}

Although the value of the exponent $\alpha_w$ for wealth 
distribution depends on the parameters of the system, 
the behaviors of the $\alpha>2$ power-law scaling for return 
distribution can exist for different parameters (e.g., coupling
parameter $a$ and random factor $\lambda$) corresponding to 
$1<\alpha_w<2$. When increasing the value of $a$, one 
still obtains the $\alpha>2$ power-law behaviors with larger 
$\alpha$ value and longer extension for the same $N$.
In Fig.\ \ref{figr500} $N=500$ and $a=0.0002$ (with the
corresponding $\alpha_w$ about $1.5$), we have 
$\alpha\sim 2.7$ but with a rather limit range of power
law regime, while for larger $a=0.0003$ as well as the same
$N$ and range of $\lambda$ (with $\alpha_w$ about 
$1.7$) $\alpha=3.1$ is obtained, and more importantly,
the power law regime is much longer (close to $3$ orders
of magnitude), as shown in Fig.\ \ref{figra23}.

\section{Properties of correlations: Correlated walk steps 
and volatility}

These results of $\alpha>2$ are nontrivial, and different
from the expectation that the wealth $w_i$ and the
fluctuation $r$ should have tails obeying the similar power 
law, since a random walk similar to Eq.\ (\ref{eqwalk}) 
with steps of sizes dominated by L\'{e}vy-like 
distribution with finite variance still leads to 
a behavior within the stable L\'{e}vy regime before the 
crossover to a Gaussian process \cite{Mantegna94}.
However, this expectation is only valid for the random
walk with statistically independent step sizes, while for
the multiplicative process (\ref{eqwi}), there are 
cross-correlations between the relative wealths $w_i(s)/W(s)$
comprising the returns in Eq.\ (\ref{eqwalk}) at each 
micro-step $s$. This can be verified by studying the
properties of time series $w_i(t)/W(t)$, where $w_i$
refers to the newly updated wealth at each time $t$.
As expected, the probability distribution of the 
walk process of Eq.\ (\ref{eqwalk}) is almost indistinguishable
from that of return $r(\tau)$ shown in Figs.\ \ref{figr500},
\ref{figr5000}, and \ref{figra23}. However, if randomizing the 
time series of $w_i(t)/W(t)$, which corresponds to eliminate 
all the possible correlations but keep the distribution form 
of step sizes, and then calculating the distribution
of $R(\tau)$ due to Eq.\ (\ref{eqwalk}), one obtains the 
completely different results, that is, the convergence to
Gaussian is faster and only the power law regions with
L\'{e}vy-like exponent $\alpha<2$ can be found, similar to the 
process of ordinary truncated L\'{e}vy flight \cite{Mantegna94}.

\begin{figure}
\centerline{
\resizebox{0.75\textwidth}{!}{%
  \includegraphics{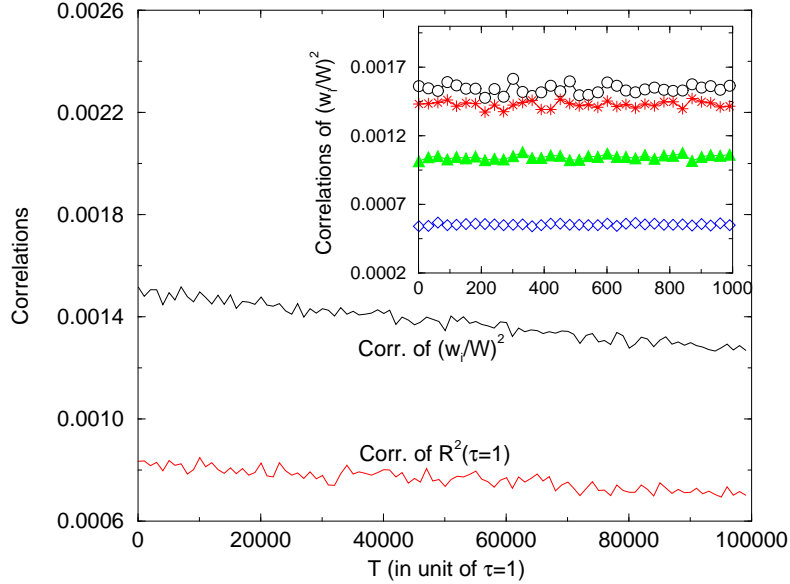}}}
\caption{Correlations of $(w_i/W)^2$ and $R^2(\tau=1)$ for $N=500$,
$a=0.0002$, and $\lambda$ in the range ($0.95$,$1.05$) (averaged
over $1100$ runs). Inset: Correlations of $(w_i/W)^2$ (averaged 
over $1000$ runs) for $N=500$ and different $a$ values and $\lambda$ 
ranges: $a=0.0008$ and $\lambda\in(0.9,1.1)$ (circles), $a=0.0002$ and 
$\lambda\in(0.95,1.05)$ (stars), and $a=5\times 10^{-5}$ and 
$\lambda\in(0.98,1.02)$ (triangles), which all correspond to the same 
exponent $\alpha_w=1.5$ of wealth distribution, as well as very small
$a=2\times 10^{-6}$ and $\lambda\in(0.995,1.005)$ (diamonds).}
\label{figcorrwi}
\end{figure}

To directly calculate the cross-correlations between
the relative updated wealths $w_i(t)/W(t)$, we use the
autocorrelation function of
\begin{equation}
Corr(T)=\frac{\langle x(t)x(t+T) \rangle - 
\langle x(t) \rangle \langle x(t+T) \rangle}
{\langle x^2(t) \rangle - \langle x(t) \rangle^2},
\label{eqcorr}
\end{equation}
for some variable $x(t)$. Although there is no correlation of 
relative wealths ($x=w_i/W$ in (\ref{eqcorr})), the nonzero
and positive correlation of the square $(w_i/W)^2$ is
found with long persistence, as shown in Fig.\ \ref{figcorrwi}, 
verifying the existence of the 
coupling between step sizes of the walk (\ref{eqwalk}) for 
this multiplicative system. These correlations are attributed 
to the social term $a \bar{w}$ introduced in our system 
(\ref{eqwi}), supplying the coupling between different wealths. 
A larger value of $a$ leads to an enhancement of 
cross-correlations, and then of $\alpha>2$ behavior, including 
the value of $\alpha$ and the extension of power-law region
(see Fig.\  \ref{figra23}). 
Even when keeping the value of $\alpha_w$ for wealth, the 
correlation of $(w_i/W)^2$ increases with larger $a$. 
In the inset of Fig.\ \ref{figcorrwi}, different values of
coupling parameter $a$ and different range of $\lambda$ are
chosen so that the corresponding $\alpha_w$ keeps the same 
value $1.5$, and from bottom to top the correlations increase 
with the increasing values of $a$.

\begin{figure}
\centerline{
\resizebox{0.75\textwidth}{!}{%
  \includegraphics{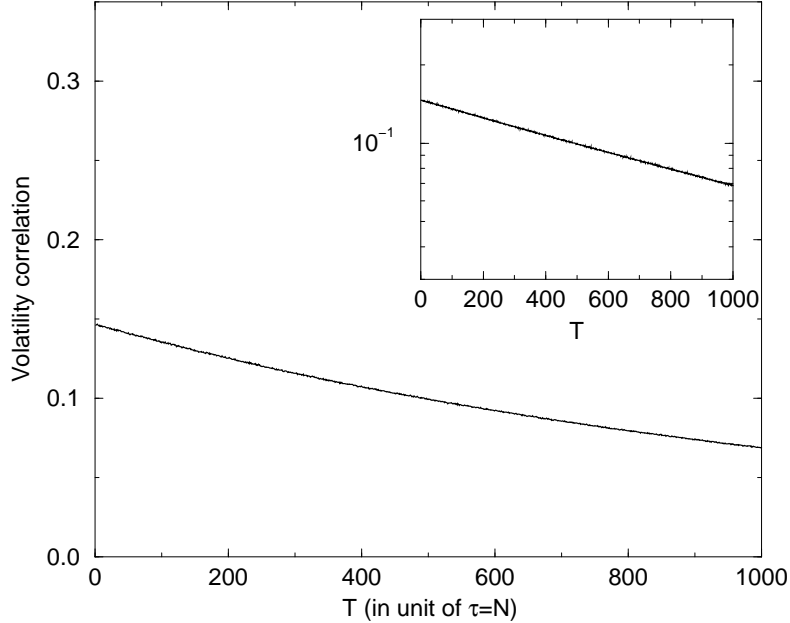}}}
\caption{Autocorrelations of $R^2(\tau=N)$ (volatility correlations,
averaged over $1500$ runs), 
for $N=500$, $a=0.0002$, and the uniform distribution of 
$\lambda$ in the range ($0.95$,$1.05$). The autocorrelation of 
$R(\tau)$, not shown, are on the scale of this figure just a flat 
line at high zero. Inset: Semi-log plot of volatility correlation.}
\label{figcorr}
\end{figure}

As the consequence of cross-correlations between 
$w_i/W$, the volatility correlations for $R(\tau)$ (which 
is the weight-sum of $w_i/W$ shown in Eq.\ (\ref{eqwalk}) 
and equivalent to return $r(\tau)$), defined as the 
autocorrelation of square of returns, i.e., $x=R^2(\tau)$
in Eq.\ (\ref{eqcorr}), exhibit the long-range persistence, 
as shown in Fig.\ \ref{figcorrwi} for $\tau=1$ (lower curve). 
Fig.\ \ref{figcorr} gives the results of volatility 
correlations for $\tau=N=500$, which exhibits the slow 
decaying and seems to be exponential-type (inset of Fig.\ 
\ref{figcorr}) rather than the power law decay found in real 
market data \cite{Louzoun01}. For correlation of return $R$ 
the value is around 0, consistent with empirical observations 
\cite{Stanley-book,Gopikrishnan99,Huang00}. This property
of volatility clustering can be also reproduced in some 
microscopic models, but there an extra dynamics like diffusion 
of traders or feedback between price change and trading activity 
is needed \cite{Stauffer}, while in this asynchronous system
it is obtained intrinsically.

\section{Discussions and conclusions}

It is interesting to relate the scales $\tau$ and $N$ here
with that of the real markets. The more natural time unit
of measurement is $\tau=N$, which corresponds to average
updating interval of stocks or agents. To describe the 
properties of market index, the relevant 
degrees of freedom $N$ could be interpreted as the major 
companies capitalization in the market, while for individual 
stocks $N$ may correspond to the important agents 
dominating the stock price. Thus, the somewhat different
behavior (value of $\alpha$, exponential cut-off, etc.)
found in different markets and stocks may be attributed
to different effective $N$. Moreover, according to our
results in Sec.\ \ref{sec3} the effective $N$ is not
very large (that is, the finite size effect is significant) 
in real market, indicating a phenomenon that is also found
in some microscopic models \cite{Sornette99}: It is a limited
number of important players, not millions of small traders,
who determine the market.

In summary, we have shown that the stochastic multiplicative
model (\ref{eqwi}) reproduces two different power law behaviors
found in reality for the wealth distribution (\ref{eqPw}) and
the return distribution (\ref{eqPr}), as well as the volatility
clustering. The dynamics of returns is described by a walk with 
steps of sizes obeying a truncated L\'{e}vy-like distribution, 
and in particular, having cross correlations. These 
cross-correlations between relative updated wealths $w_i(t)/W(t)$ 
are expected to be the origin of the $\alpha>2$ behavior in the 
tail distribution and the long-range volatility correlations
of returns, and can be attributed to the coupling term in 
multiplicative system (\ref{eqwi}).

\section*{Acknowledgements}
We thank Dietrich Stauffer, Ofer Biham, and Ofer Malcai
for very helpful discussions and comments.


\begin{thebibliography}{}

\bibitem{Kesten73} H. Kesten, Acta Math. 131 (1973) 207.

\bibitem{Redner90} S. Redner, Am. J. Phys. 58 (1990) 267.

\bibitem{Solomon96} M. Levy and S. Solomon, Int. J. Mod. Phys. C
7 (1996) 595; S. Solomon and M. Levy, {\it ibid.} 7 (1996) 745.

\bibitem{Solomon98} S. Solomon, in: A.-P. Refenes, A.N. Burgess, 
and J.E. Moody (Eds), Decision Technologies for
Computational Finance, Kluwer Academic Publishers, 1998.

\bibitem{Sornette} D. Sornette, Physica A 250 (1998) 295;
D. Sornette and R. Cont, J. Phys. I 7 (1997) 431.

\bibitem{Marsili98} M. Marsili, S. Maslov, and Y.C. Zhang,
Physica A 253 (1998) 403.

\bibitem{Bouchaud00} J.P. Bouchaud and M. M\'ezard, Physica A
282 (2000) 536.

\bibitem{Pareto} V. Pareto, Cours d'Economique Politique, Vol. 2,
Macmillan, Paris, 1897.

\bibitem{Atkinson78} A.B. Atkinson and A.J. Harrison,
Distribution of Total Wealth in Britain, Cambridge University
Press, Cambridge, 1978.

\bibitem{Ijiri77} Y. Ijiri and H.A. Simon, Skew Distributions
and the Sizes of Business Firms, North-Holland, Amsterdam, 1977.

\bibitem{Levy37} P. L\'{e}vy, Theorie de l'Addition des 
Variables Aleatoires, Gauthier-Villiers, Paris, 1937.

\bibitem{Stanley-book} R. Mantegna and H.E. Stanley, An 
Introduction to Econophysics: Correlations and Complexity in 
Finance, Cambridge University Press, Cambridge, 1999.

\bibitem{Bouchaud-book} J.-P. Bouchaud and M. Potters, Theory 
of Financial Risk, Cambridge University Press, Cambridge, 2000.

\bibitem{Lux96} T. Lux, Appl. Financial Economics 6 (1996) 463.

\bibitem{Gopikrishnan99} P. Gopikrishnan, V. Plerou, L.A.N. Amaral, 
M. Meyer, and H.E. Stanley, Phys. Rev. E 60 (1999) 5305.

\bibitem{Plerou99} V. Plerou, P. Gopikrishnan, L.A.N. Amaral, 
M. Meyer, and H.E. Stanley, Phys. Rev. E 60 (1999) 6519.

\bibitem{Huang00} Z.F. Huang, Physica A 287 (2000) 405; 
L.H. Tang and Z.F. Huang, Physica A 288 (2000) 444.

\bibitem{Skjeltorp00} J. Skjeltorp, Physica A 283 (2000) 486.

\bibitem{Masoliver00} J. Masoliver, M. Montero, and J.M. Porr\`{a},
Physica A 283 (2000) 559.

\bibitem{Solomon01} S. Solomon and P. Richmond, e-print,
cond-mat/0102423, to be published in Physica A.

\bibitem{Biham-Malcai} O. Biham, O. Malcai, M. Levy, and S. Solomon,
Phys. Rev. E 58 (1998) 1352; O. Malcai, O. Biham, and 
S. Solomon, Phys. Rev. E 60 (1999) 1299.

\bibitem{Huang-Solomon01} Z.F. Huang and S. Solomon, 
Eur. Phys. J. B 20 (2001) 601.

\bibitem{Biham01} O. Biham, Z.F. Huang, O. Malcai, and S. Solomon,
Phys. Rev. E 64 (2001) 026101.

\bibitem{note2} Note that at each time step $s$ the label $i$ of
the updated wealth $w_i$ changes and is actually $i(s)$.

\bibitem{Huang01} Z.F. Huang and S. Solomon, Physica A 294
(2001) 503.

\bibitem{Mantegna94} R.N. Mantegna, H.E. Stanley, Phys. Rev. Lett.
73 (1994) 2946; I. Koponen, Phys. Rev. E 52 (1995) 1197.

\bibitem{Stauffer99} D. Stauffer and D. Sornette, Physica A
271 (1999) 496; F. Castiglione, R.B. Pandey, and
D. Stauffer, Physica A 289 (2001) 223.

\bibitem{Louzoun01} The power-law correlations can be reproduced
by assuming the variance of $\lambda$ as a function of market
volatility, as shown in: Y. Louzoun and S. Solomon, to be published 
in Physica A.

\bibitem{Stauffer} D. Stauffer, P.M.C. de Oliveira, and A.T.
Bernardes, Int. J. Theor. Appl. Finance 2 (1999) 83;
D. Stauffer and N. Jan, Physica A 277 (2000) 215;
Z.F. Huang, Eur. Phys. J. B 16 (2000) 379.

\bibitem{Sornette99} D. Sornette, D. Stauffer, and H. Takayasu,
e-print, cond-mat/9909439.

\end{thebibliography}
\end{document}